\documentclass{ws-procs11x85}
\def\lapprox{\lower .7ex\hbox{$\;\stackrel{\textstyle <}{\sim}\;$}}
\def\gapprox{\lower .7ex\hbox{$\;\stackrel{\textstyle >}{\sim}\;$}}

\makeindex
\begin{document}

\title{QCD: Theoretical Developments}

\author{Thomas Gehrmann}

\address{Institut f\"ur Theoretische Physik, Universit\"at Z\"urich,
Winterthurerstra\ss{}e 190, CH-8057 Z\"urich, Switzerland}


\twocolumn[\maketitle\abstract{ 
I review recent theoretical advances in quantum chromodynamics. Particular 
emphasis is put on developments related to the precise prediction and 
interpretation of experimental data from present and future high energy 
colliders.}]

\baselineskip=13.07pt
\section{Introduction}
Quantum Chromodynamics (QCD) is well established as theory of strong 
interactions through a large number of experimental verifications. 
The era of `testing QCD' is clearly finished,
and QCD today is becoming precision physics. The next generation of 
high energy collider experiments are all performed at hadron 
colliders, 
where (in contrast to LEP and SLC)  QCD is 
ubiquitous. Any precision measurement (strong coupling constant, quark masses, 
electroweak parameters, parton distributions) 
at the Tevatron and the LHC, as well as any prediction of new 
physics effects and their backgrounds, relies on the understanding of 
QCD effects on the observable under consideration. 

The derivation of precise QCD predictions for collider observables 
poses several  theoretical and computational challenges.
The most important  
challenge is the fact that QCD describes quarks and gluons, while 
experiments observe hadrons. This mismatch is either accounted for by 
a description of the parton to hadron 
transition through fragmentation functions or by defining sufficiently 
inclusive final state observables, such as jets. 
Moreover, the strong coupling 
constant is considerably larger than the electromagnetic coupling constant 
at scales typically probed at colliders: $\alpha_s(M_Z) \simeq 15\, 
\alpha_{{\rm em}} (M_Z)$, resulting in a slower convergence of the 
perturbative expansion. As a consequence, a precise description of 
QCD observables (precise means here that the theoretical 
uncertainty becomes similar to the achieved or projected experimental errors)
 is obtained only by including higher order corrections, often 
requiring beyond the next-to-leading order. The largeness of the strong 
coupling constant also implies that multiparticle final states 
are rather frequent. 
Finally, many collider observables involve largely different scales, such as 
quark masses, transverse momenta and vector boson masses. These give rise 
to potentially large logarithms, which might spoil the convergence of the 
perturbative series and need to be resummed to all orders. 

In this talk, I shall try to highlight recent theoretical progress towards 
precision QCD at colliders, focusing on heavy quark production in 
Section~\ref{sec:hq}, on jets and multiparton final states in 
Section~\ref{sec:jets}, on photons in 
Section~\ref{sec:photons} and electroweak bosons in 
Section~\ref{sec:vecbos}. Finally, a summary of the current state-of-the-art
and of yet open issues is given in Section~\ref{sec:conc}.

\section{Heavy Quarks}
\label{sec:hq}
Heavy quark production is one of the main topics investigated at 
high energy collider experiments.  Heavy quarks are of particular interest 
to probe the flavour sector of the standard model, which is less well 
tested than the gauge sector. 
Also, many approaches to physics beyond the standard model, often 
related to electroweak symmetry breaking and mass generation, predict 
new effects to be most pronounced in observables involving heavy quarks. 
\begin{figure}
\center
\psfig{figure=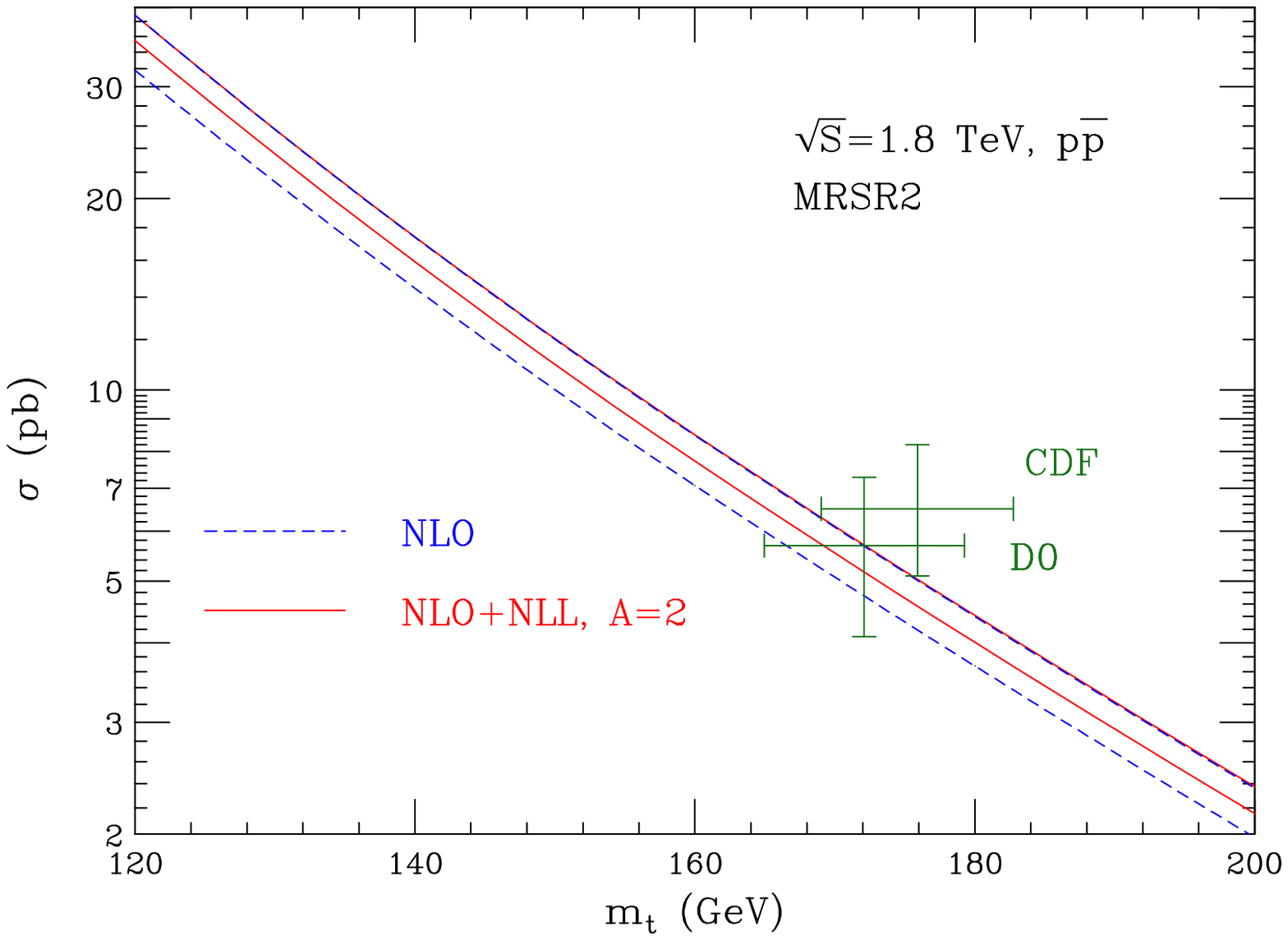,width=7.0truecm}
\psfig{figure=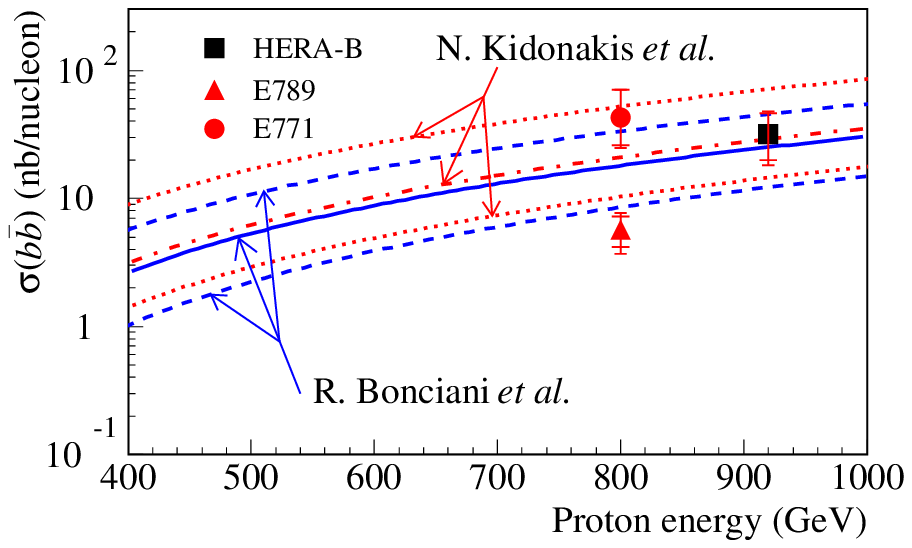,width=8.0truecm}
\caption{Total cross sections for $t\bar t$ at the 
Tevatron\protect\cite{hqnll} and 
$b\bar b$ at HERA-B\protect\cite{hqherab}.}
\label{fig:hqtotal}
\end{figure}

\subsection{Total Cross Sections}
The total cross sections for the production of heavy quarks can be
computed reliably within perturbation theory. The current state-of-the-art is 
a next-to-leading order (NLO) calculation\cite{hqnlo}, which is
further improved by summing large logarithms due to soft gluon
emission up to the next-to-leading  (NLL)\cite{hqnll}
and next-to-next-to-leading logarithmic (NNLL)\cite{hqnnll}
level. As can be seen from Figure~\ref{fig:hqtotal},
these predictions are in good agreement with experimental data on the
total $t\bar t$ cross section at the Tevatron\cite{hqcoll} 
and the total $b\bar b$ 
cross section at HERA-B\cite{hqherab}
(which both actually refer to similar kinematical 
values of $m_Q/\sqrt{s}$). The theoretical uncertainty on the
prediction for HERA-B is larger for two reasons: the larger value of 
the strong
coupling at $m_b$ than at $m_t$ and the dominance of $gg$ initial
states in $pN$ collisions (HERA-B) compared to $q\bar q$ dominance in 
$p\bar p$ collisions (Tevatron). The effects due to soft gluon resummation 
turn out to be moderate, but do yield a significant decrease in the 
uncertainty of the theoretical prediction. Further uncertainties on the 
prediction of the $t\bar t$ cross section come from the 
parton distribution functions\cite{tterr}.

\subsection{Transverse Momentum Distributions}
Differential distributions of hadrons containing $b$ quarks 
measured in hadron-hadron, photon-hadron or photon-photon collisions 
have been in apparent discrepancy with theoretical predictions for 
quite some time. The spectrum of $B^\pm$ hadrons measured at 
CDF\cite{cdftrmom} 
is one of the most recent examples for this discrepancy.

The theoretical prediction for $B$ meson production involves a convolution 
of the hard matrix element for heavy quark production in parton-parton 
scattering with initial parton distributions and final state fragmentation 
functions describing the non-perturbative transition from a $b$ quark to 
a $B$ hadron. It is in particular the latter which 
is suspected to  account for the 
discrepancy between theoretical prediction and experimental data, 
especially since it has been observed\cite{d0bjet} 
that the transverse momentum 
distribution of $b$-tagged jets\cite{frixione} (which 
has little sensitivity to fragmentation functions) is in much
better agreement with theoretical predictions. 

The definition of heavy quark
fragmentation functions is not free from ambiguities, since 
some aspects of these functions are actually calculable in 
perturbation theory\cite{mele}. In extracting these fragmentation functions 
from data on $B$ hadron production in $e^+e^-$ collisions, several choices 
are made, related to the order of perturbation theory, the incorporation of 
mass effects in the matrix elements, the resummation of potentially large
perturbative terms, the inclusion of power corrections\cite{cagar}, 
the correction of data for parton showers or the 
parametric form of the ansatz used in the determination. 
Unfortunately, the sensitivity of the fragmentation function
on the assumptions 
used in the extraction from $e^+e^-$ spectra is often overlooked when 
using this fragmentation function to compute heavy hadron spectra 
at colliders. 

Recently, an approach
incorporating quark mass effects, perturbatively calculable components of 
the heavy quark fragmentation function\cite{mele} and resummation of 
large logarithms up to the next-to-leading logarithmic level has been put 
forward with the fixed-order next-to-leading log
(FONLL) scheme\cite{cacciari1}.
This approach requires only a small, genuinely non-perturbative 
component of the fragmentation function to be fitted to $e^+e^-$ data. 
In order to expose the information content actually relevant to 
heavy hadron spectra at hadron colliders, this fit is done in moment 
space. 

\begin{figure}[tb]
\begin{center}
\epsfig{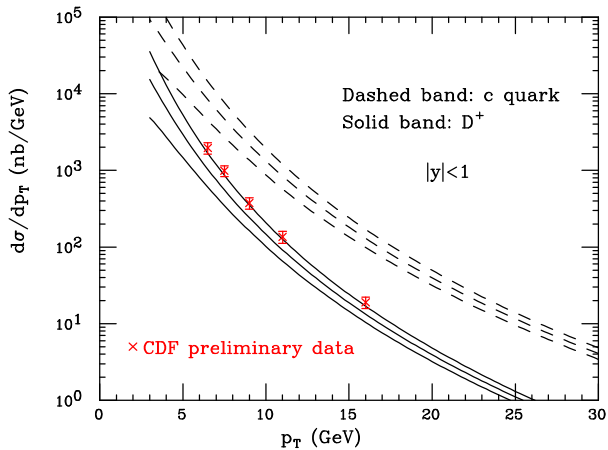}
\end{center}
\caption{Transverse momentum spectrum of $D^+$ hadrons at 
CDF, compared 
to calculation using FONLL fragmentation functions\protect\cite{cncharm}.}
\label{fig:hqnll}
\end{figure}

In view of new data from ALEPH\cite{lephq}, a phenomenological study 
of $B$ hadron production at colliders based on the FONLL 
scheme was performed\cite{cacciari}. It 
was shown that the consistent treatment of the fragmentation function 
in extraction and prediction reduced the discrepancy between 
data and theoretical prediction considerably. 
The theoretical prediction is however still falling somewhat  
short of the experimental data, which is probably due to currently 
uncalculated corrections beyond NLO. More recently, the same framework
was applied to charmed hadron production at hadron 
colliders\cite{cncharm}. In this case, one also observes that the 
experimental data\cite{charmdata}
 exceed the theoretical prediction, Figure~\ref{fig:hqnll}, 
although the effect is 
less pronounced than for bottom hadron production. 
Comparison of massless\cite{kkcharm} and massive\cite{cncharm} 
calculations for hadronic charm production\cite{charmdata} indicates 
only moderate mass effects, which are smaller than the theoretical 
uncertainty. 

Many collider experiments also report an excess in the 
$b$ quark production spectra. In interpreting these data, it must 
always be kept in mind that it is not $b$ quark but $B$ hadron production 
which is observed in the experiment. Information on $b$ quark production 
is only inferred from these data using some model for the 
heavy quark fragmentation. As discussed above, there are numerous ambiguities, 
which can yield inconsistent predictions if not implemented consistently.
In view of the rather sizable effects due to a consistent 
treatment of the fragmentation function observed on the $B$ hadron spectra 
at CDF, it might be that the data sets on $b$ quark spectra have to be 
reanalyzed incorporating the new experimental and 
theoretical information on
the $b$ quark fragmentation functions in a consistent manner.

\section{Jets and Multiparticle Production}
\label{sec:jets}
Hadronic jets at large transverse momenta are produced very copiously at 
colliders. Final states with a small number of jets are measured to 
very high experimental accuracy, such that they can be used for 
precision measurements of the strong coupling constant and of 
parton distribution functions. For these measurements, the uncertainty 
on the theoretical prediction is often the dominant source of error, and one 
would consequently like to have theoretical calculations to be more accurate, 
which implies in general an extension towards higher perturbative orders. 
Multiparton final states, involving a 
large number of jets, can on the other hand mimic final state signatures 
induced by physics beyond the standard model, thus forming an irreducible 
background to searches. For these, QCD predictions serve as a guidance to 
devise search strategies, and one demands QCD to yield a description of the 
full hadronic final state. 

\subsection{Leading order calculations}
Multiparton final states are described using leading order 
QCD predictions, implemented in flexible multiparton matrix element 
generators. These programs evaluate the scattering amplitudes using 
efficient representations of helicity amplitudes or fully numerically 
from the interaction Lagrangian. Examples of these codes are 
VECBOS\cite{vecbos}, COMPHEP\cite{comphep}, 
MADGRAPH\cite{madgraph},~GRACE\cite{grace},~HELAC\cite{helac}, 
ALPHGEN\cite{alphgen}
and AMEGIC++\cite{amegic}. Using these, the 
computation of $2\to 8$ reactions is 
feasible on current computers. These programs are then combined with 
automatic integration over multiparticle phase spacs, using 
for example RAMBO\cite{rambo}, PHEGAS\cite{phegas} or 
MADEVENT\cite{madevent}. 
Most programs can be interfaced to hadronization models using 
standard interfaces\cite{lh01}. 

Matrix element calculations accurately include large angle single 
gluon radiation. At small angles from the emitting particles, one does 
however encounter multiple gluon radiation, which can be  accounted for
by parton showers. Recently, a generic procedure was devised to 
combine both descriptions for multiparton final states 
in a modified matrix element plus vetoed 
parton shower\cite{ckkw}. 

Including these developments, leading order QCD provides the basis of 
Monte Carlo event generators. However, its predictions contain large (and 
non-quantifiable) errors due to the setting of renormalization and 
factorization scales. Leading order
QCD is therefore good tool to estimate relative magnitudes of
processes and to design searches. Once precision is required (e.g.\ to
identify a discovery with a particular model), it is not suff\/icient.

\subsection{Next-to-leading order calculations}
Including next-to-leading order QCD corrections improves the theoretical 
predictions in numerous ways by reducing the renormalization scale 
uncertainty, providing reliable normalizations of cross sections, and 
reliable error estimates. Moreover, NLO is the first order 
where differences between jet algorithms show up. In contrast to 
leading order, there is no generic procedure for doing NLO calculations, 
such that each new process under consideration implies a 
completely new calculation. 

\begin{figure}[tb]
\begin{center}
\epsfig{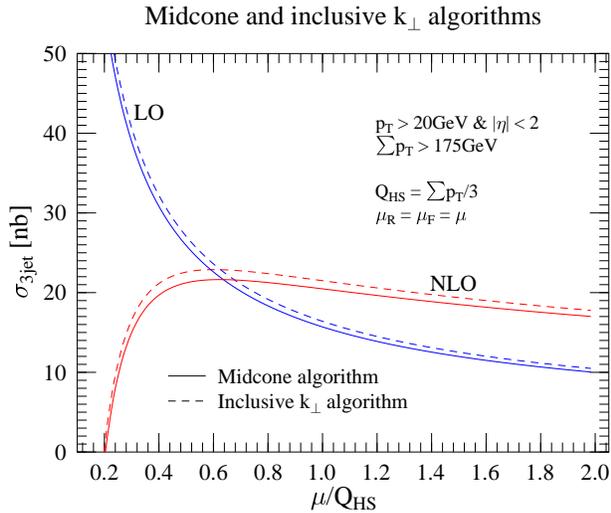} 
\end{center}
\caption{Dependence of the NLO prediction for the $3j$ cross section at the 
Tevatron on renormalization and factorization scale\protect\cite{nagy}.}
\label{fig:nlojet}
\end{figure}

For hadron colliders, NLO results are available 
for all relevant $2\to 2$ reactions; $2\to 3$ reactions are the current 
frontier. A number of $2\to 3$ results (each 
involving  several man-years of work) became available recently:
$pp\to V+ 2j$\cite{mcfm}, $ep\to (3+1)j$\cite{nagyep}, $pp \to 3j$\cite{nagy},
$pp\to \gamma\gamma + j$\cite{ggjet},
$pp\to t\bar t H$\cite{tth} and 
the vector boson fusion processes $pp \to H +2j$\cite{oz}, 
$pp \to V +2j$\cite{oz2}. 
Some of the features of NLO calculations, such as the improved scale 
dependence and the differences between jet algorithms are illustrated in 
Figure~\ref{fig:nlojet}.

To overcome the large amount of work required for each NLO calculation, 
efforts are under way towards their automatization. The NLO calculation 
for an $n$ parton reactions contains 
the one-loop $n$   parton matrix elements, the tree level $n+1$ parton 
matrix elements and a procedure to extract the infrared singularities from 
both and to combine them. While this procedure has been automatized 
for the tree level real radiation matrix elements
long ago\cite{ggcs}, there is at present no 
automatic procedure to compute one-loop integrals. 
Very recently several algorithms were proposed, including the 
analytic reduction of hexagon integrals\cite{hex}, a subtraction 
formalism for virtual corrections\cite{virtnagy} and the numerical 
evaluation of hexagon integrals\cite{hexnum}.

\begin{figure}[tb]
\begin{center}
\epsfig{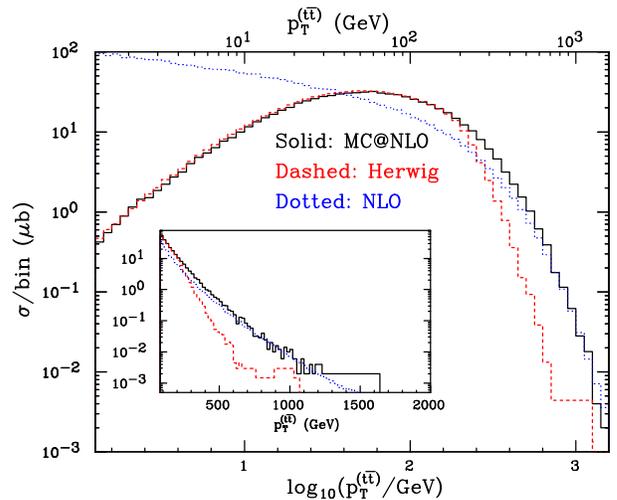} 
\end{center}
\caption{Transverse momentum distribution of top quark pairs 
at the LHC as predicted by MC@NLO\protect\cite{mcnlott}.}
\label{fig:mcnlo}
\end{figure}

Another important development is the combination of NLO calculations with 
parton showers, as realized in the MC@NLO approach\cite{mcnlo}. This 
approach introduces a modified NLO subtraction method, where both real and 
virtual contributions become initial conditions for the parton shower. 
In this, hard radiation is accurately described by the NLO matrix element, 
while multiple soft radiation is accounted for by the parton shower; a
double counting of contributions is avoided. So far, this formalism 
has been applied to $VV$, $b\bar b$ and $t\bar t$ production at hadron 
colliders. Figure~\ref{fig:mcnlo} illustrates that 
MC@NLO\cite{mcnlott} smoothly 
connects the kinematic region dominated by multiple radiation at small 
transverse momenta to the region controlled by single hard radiation at 
large transverse momenta. 

\subsection{Next-to-next-to-leading order calculations}
\label{sec:jetnnlo}
\begin{figure}[t]
\begin{center}
\epsfig{file=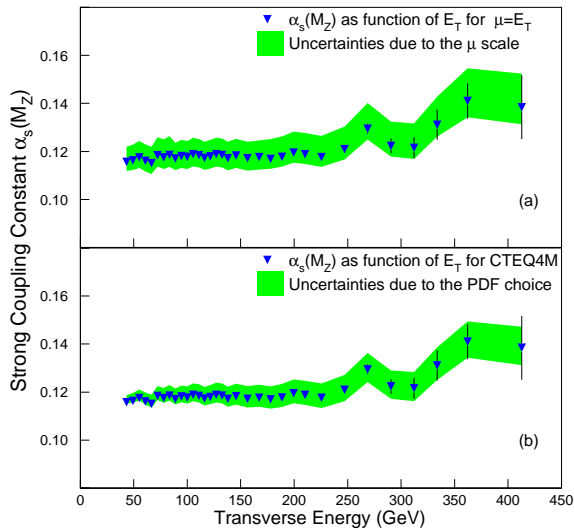,width=8cm} 
\end{center}
\caption{Errors on extraction of $\alpha_s$ from single jet inclusive 
cross section at CDF~\protect\cite{cdfjet}.}
\label{fig:als}
\end{figure}

Despite the evidently good agreement of NLO QCD with experimental data
on jet production rates, predictions to this
order are insufficient for many applications. For example, if one uses
data on the single jet inclusive cross section\cite{cdfjet}
compared to the NLO theoretical prediction\cite{jetrad,dyrad}
to determine the strong
coupling constant $\alpha_s$, it turns out that the dominant source of 
error on this extraction is due to unknown higher order corrections. 
Given that the theoretical prediction to infinite order in
perturbation theory should be independent of the choice of
renormalization and factorization scale, this error can be estimated
from the variation of the extracted $\alpha_s$ under variation of
these scales, as seen in Figure~\ref{fig:als}. As a result, CDF find 
from their Run I data
\begin{eqnarray*}
\alpha_s (M_Z)
&=& 0.1178 \pm 0.0001 (\mbox{stat})^{+0.0081}_{-0.0095}
(\mbox{sys})\\
&&\;^{+0.0071}_{-0.0047}(\mbox{scale})\pm 0.0059 (\mbox{pdf}).
\end{eqnarray*}
It can be seen that the statistical error is already negligible;
improvements in the systematic error can be anticipated in the near
future. To lower the theoretical error, it is mandatory to compute 
next-to-next-to-leading order (NNLO) corrections to the single jet
inclusive cross section.

A similar picture is true in $e^+e^-$ annihilation into three jets and 
deep inelastic $(2+1)$ jet production, where the error on the 
extraction of $\alpha_s$ from experimentally measured jet shape 
observables\cite{icheplong} is completely dominated by the 
theoretical uncertainty inherent in the NLO QCD calculations. 

Besides lowering the theoretical error, there is  a number of other 
reasons to go beyond NLO in the description of jet
observables. While jets at NLO are modeled theoretically
by at most two partons, NNLO allows up to three partons in a single
jet, thus improving the matching of experimental and theoretical 
jet definitions and resolving the internal jet structure. At hadron
colliders, NNLO does also account for double initial state radiation,
thus providing a perturbative description for the transverse momentum
of the hard final state. Finally, including jet data in a global NNLO
fit of parton distribution functions, one anticipates a lower error on
the prediction of benchmark processes at colliders.

The calculation of jet observables at NNLO requires a number of
different ingredients. To compute the corrections to an $n$-jet
observable, one needs the two-loop $n$ parton matrix elements, 
the one-loop $n+1$ parton matrix elements and the tree level $n+2$
parton matrix elements. Since the latter two contain infrared
singularities due to one or two partons becoming theoretically
unresolved (soft or collinear), one needs to find one- and
two-particle subtraction terms, which account for these singularities 
in the matrix elements, and are sufficiently simple to be integrated
analytically over the unresolved phase space. One-particle
subtraction at tree level is well understood from NLO 
calculations\cite{ggcs} and general algorithms are available for 
one-particle subtraction at one loop\cite{onel}, in a 
form that could recently be integrated analytically\cite{wz1}. Tree level
two-particle subtraction terms have been extensively studied in the 
literature\cite{twot},
their integration over the unresolved phase space was up to now made
only in one particular infrared subtraction scheme in the
calculation of higher order corrections to the photon-plus-one-jet
rate in $e^+e^-$ annihilation\cite{ggam}. The same techniques
(and the same scheme) were used 
very recently in the rederivation of the time-like gluon-to-gluon
splitting function from splitting amplitudes\cite{dak}.
A general two-particle subtraction
procedure is still lacking at the moment, although progress on this is 
anticipated in the near future. 

Concerning virtual two-loop corrections to jet-observables related to 
$2\to 2$ scattering and $1\to 3$ decay processes, enormous progress
has been made in the past years. Much of this progress is due to 
several technical developments concerning the evaluation of two-loop
multi-leg integrals. Using iterative  
algorithms\cite{laporta}, one can  reduce the
large number of two-loop integrals by means of
integration-by-parts\cite{chet} and Lorentz invariance\cite{gr}
identities to a small number of master integrals. The master
integrals relevant to two-loop jet physics are two-loop four-point
functions with all legs on-shell\cite{onshell} or one leg
off-shell\cite{mi},
which were computed using explicit integration or implicitly from
their differential equations\cite{gr}. 

Combing the reduction scheme with the master integrals, it is 
straightforward to compute the two-loop matrix elements relevant to
jet observables using computer algebra\cite{radcor}.
Following this procedure, massless
two-loop matrix elements were obtained for 
Bhabha scattering\cite{m1}, parton-parton scattering into two 
partons\cite{m2}, parton-parton scattering into two 
photons\cite{m3}, as well as light-by-light
scattering\cite{m4}. Two-loop corrections were also 
computed for the off-shell process $\gamma^*\to q\bar q
g$\cite{3jme}, relevant to $e^+e^-\to 3j$. Part of these results were
already confirmed\cite{muw2} using an independent
method\cite{muw1}. Related to  $e^+e^-\to 3j$ by analytic
continuation\cite{ancont} are $(2+1)j$ production in $ep$ collisions and $V+j$
production at hadron colliders.
A strong check on all these two-loop results is provided by the
agreement of the singularity structure with predictions obtained from
an infrared factorization formula\cite{catani}.

More recently, first results were obtained for master integrals involving 
massive internal propagators, as appearing in the two-loop 
QED corrections to the $\gamma^*\to Q\bar Q$ 
vertex\cite{bonc1} or in the two-loop electroweak corrections to the 
$V \to q\bar q$ vertex\cite{bonc2}.

\section{Photons}
\label{sec:photons}
Photons and gauge bosons provide very prominent final state signatures at 
colliders.
Their study allows the precise determination of electroweak parameters 
at hadron colliders, and their final state signatures are often background
to searches, such as photon pair production to the Higgs search in the 
lower mass range. 

\subsection{Isolated Photons}
Photons produced in hadronic collisions arise essentially from two different 
sources: `direct' or `prompt' photon production via hard partonic processes
such as $qg\to q\gamma$ and $q\bar q\to g\gamma$ or through the `fragmentation'
of a hadronic jet into a single photon carrying a large fraction of the 
jet energy. The former gives rise to perturbatively calculable short-distance 
contributions whereas the latter is primarily a long distance process which 
cannot be calculated perturbatively and is described in terms of the 
quark-to-photon fragmentation function. In principle, this fragmentation 
contribution could be suppressed to a certain extent by imposing isolation 
cuts on the photon. Commonly used isolation cuts are defined by admitting only 
a maximum amount of hadronic energy in a cone of a given radius around the 
photon. An alternative procedure is the democratic clustering approach 
suggested in\cite{morgan}, which applies standard jet clustering algorithms 
to events with final state photons, treating the photon like any other hadron 
in the clustering procedure. Isolated photons are then defined to be 
photons carrying more than some large, predefined amount of the jet energy. 

Both types of 
isolation criteria infrared safe, although the matching of experimental 
and theoretical implementations of these criteria is in general far from 
trivial. It was pointed out recently\cite{cf} that cone-based isolation 
criteria fail for small cone sizes $R$ (once $\alpha_s\ln R^{-2}\sim 1$), 
since the isolated photon cross section exceeds the inclusive photon cross 
section. This problem can only be overcome by a resummation of the large 
logarithms induced by the cone size parameter.
The contribution from photon fragmentation to isolated photon 
cross sections at hadron colliders 
is sensitive (for both types of isolation criteria) on 
the photon fragmentation function at large momentum transfer, which 
has up to now been measured only at LEP\cite{aleph,opal}. 
Further information on the photon fragmentation function at large momentum 
transfer might be gained from yet unanalyzed LEP data or from 
the study of photon-plus-jet final states in deep inelastic scattering at
HERA\cite{gks}, where first data are now becoming available\cite{lemrani}.

\subsection{Photon Pairs}
One of the most promising channels for the 
discovery 
of a light Higgs boson ($m_H \lapprox 140$~GeV) at the LHC 
is based largely on the observation of the 
rare decay to two photons. To perform an accurate 
background subtraction for this observable, one requires a precise 
prediction for QCD reactions yielding di-photon final states. At first 
sight, the ${\cal O}(\alpha_s^0)$ process $q\bar q\to \gamma\gamma$
yields the leading contribution. However, due to the large gluon luminosity 
at the LHC, both $qg\to q \gamma\gamma$ $({\cal O}(\alpha_s^1))$ and 
$gg\to \gamma\gamma$ $({\cal O}(\alpha_s^2))$ 
subprocesses yield contributions 
of comparable magnitude. The NLO corrections to 
the $q\bar q$ and $qg$ subprocesses have been  known for quite some time,
these are  implemented in the flexible parton level event 
generator DIPHOX\cite{diphox}. Most recently, NLO corrections were 
also derived for the $gg$ subprocess\cite{bernschmidt}. Since the 
lowest order contribution to this process is already mediated by a quark loop, 
this calculation contains some of the features appearing in 
jet physics only at NNLO, such as two-loop 
QCD amplitudes and unresolved limits of one-loop amplitudes 
(see Section~\ref{sec:jetnnlo} above). Another important 
new result are the NLO corrections to two-photon-plus-jet 
production\cite{ggjet}, forming the  background to Higgs boson 
detection at large transverse momenta.
\begin{figure}[tb]
\begin{center}
\epsfig{file=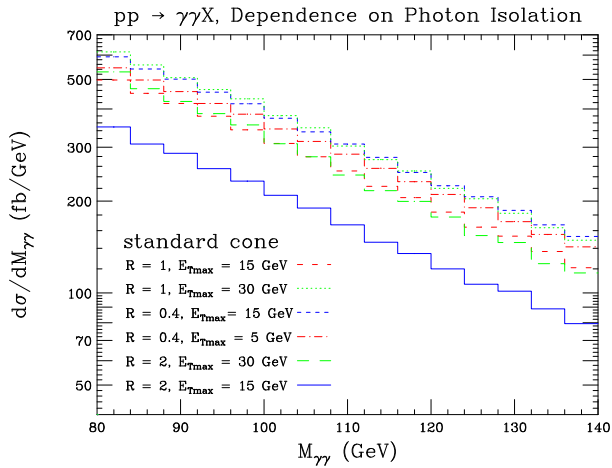,width=8cm}
\end{center}
\caption{Photon pair production 
 for different isolation criteria\protect\cite{bernschmidt}.}
\label{fig:gparton}
\end{figure}

It must be kept in mind that the di-photon cross sections are highly sensitive 
on the isolation criteria applied to the photons, Figure~\ref{fig:gparton},
with a substantial contribution arising  from photon fragmentation at large 
momentum transfers\cite{diphoxfrag}. 
Moreover, it is experimentally difficult to distinguish
photons from highly energetic neutral pions which decay into a closely 
collimated photon pair, mimicking a single photon signature.
The pion background in photon pair production has been studied 
to NLO\cite{diphoxpi} and implemented in DIPHOX, showing that in particular
the $\pi^0\gamma$ channel remains comparable to the $\gamma\gamma$ channel 
even for tight isolation criteria.

\section{Higgs and Gauge Boson Production}
\label{sec:vecbos}

The search for the Higgs boson is one of primary goals of 
present and future hadron collider experiments, where one expects the main 
production channel to be gluon fusion, mediated through a top quark loop. 
To a good approximation\cite{spirah}, 
one can use an effective gluon-gluon-Higgs 
coupling to describe this process in perturbative QCD (provided 
the leading order mass dependence is factored out explicitly).  In this 
approximation, the calculation higher order corrections to 
 inclusive Higgs production becomes very similar to the analogous
calculation for gauge boson production. 

Inclusive 
vector boson production has been computed to 
NNLO\cite{dynnlo} already more than ten years ago.
Very recently, these results have been verified for the first time 
in an independent calculation\cite{hkhiggs}, carried 
out in the context of  the derivation of 
NNLO corrections to  inclusive Higgs boson production.

\subsection{Higgs Boson}

\begin{figure}[tb]
\begin{center}
\epsfig{file=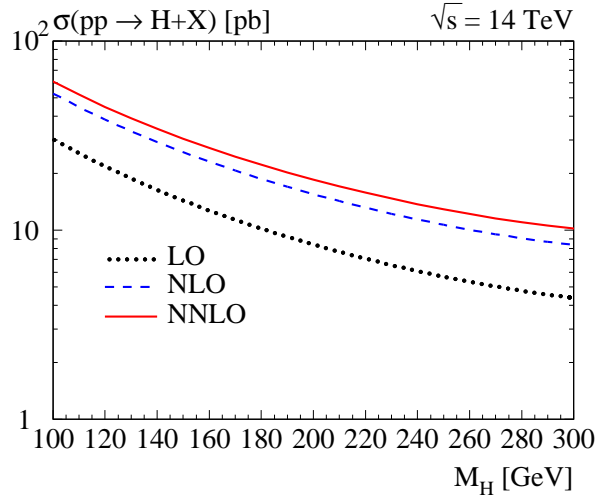,width=8cm}
\end{center}
\caption{Inclusive Higgs production at the LHC\protect\cite{hkhiggs}.}
\label{fig:higgs}
\end{figure}

The NNLO  corrections to the Higgs production cross section were 
obtained 
first in the soft/collinear approximation\cite{hggsoftcol}; shortly 
thereafter, the full coefficient functions were obtained by 
expansion around the soft limit\cite{hkhiggs},
 and  fully analytically\cite{babis} 
by extending the IBP/LI reduction method and the 
differential equation technique (see Section~\ref{sec:jetnnlo}) to
compute double real emission contributions. These results were 
confirmed independently\cite{vnhiggs} using the techniques of the 
original vector boson calculation. 
 It turned out that 
inclusion of NNLO corrections yields a sizable enhancement of the Higgs 
production cross section, Figure~\ref{fig:higgs}, 
and a reduction of the uncertainty due to 
renormalization and factorization scale. Recently, this calculation 
was further improved by the inclusion of effects due to soft gluon 
resummation\cite{catanihiggs}. Further NNLO results on inclusive
Higgs boson production involve pseudoscalar Higgs production 
through gluon fusion\cite{hp}, Higgs production in bottom quark 
fusion\cite{hbb} and Higgs-strahlung off a vector boson\cite{hz}.

\begin{figure}[tb]
\begin{center}
\epsfig{file=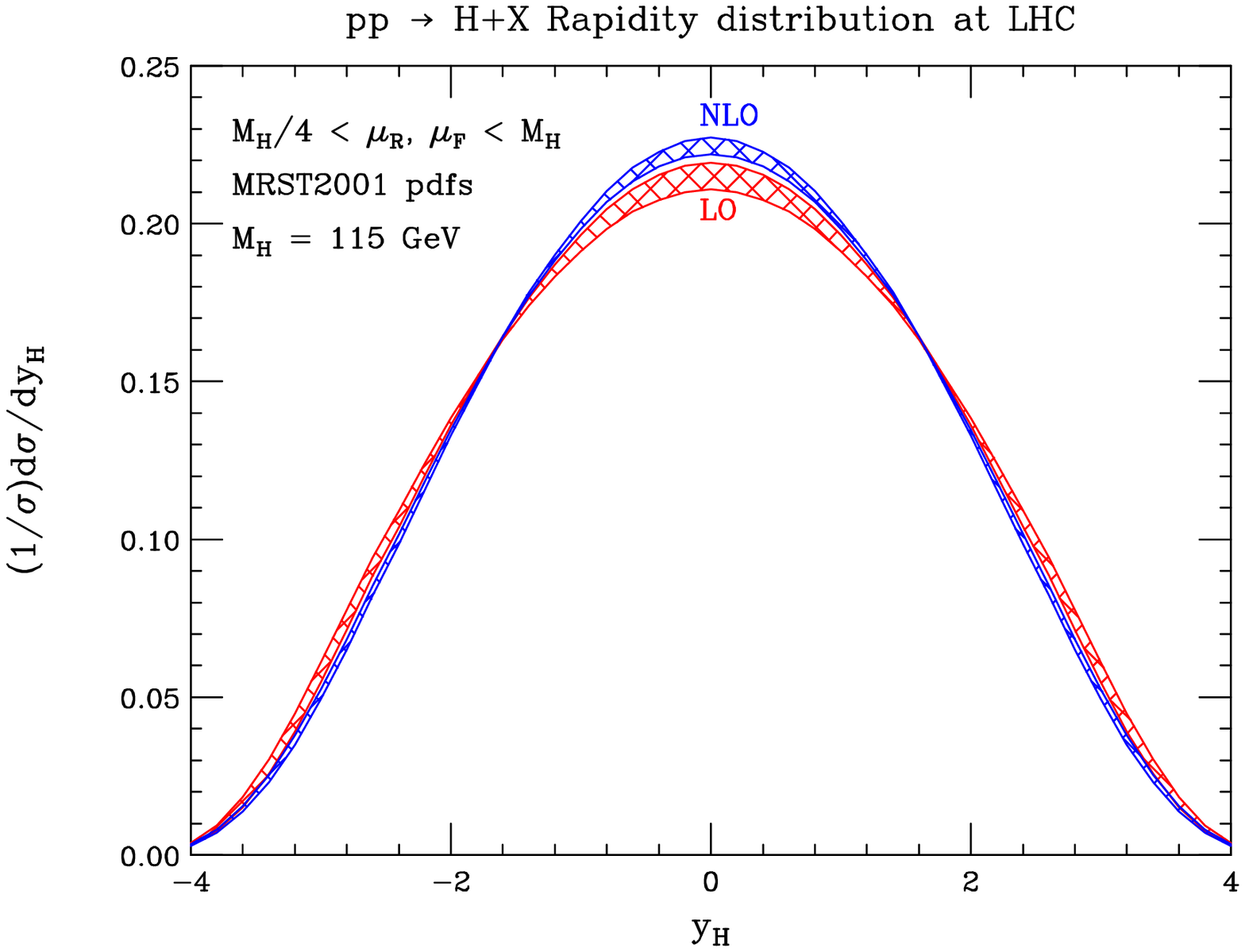,width=8cm}
\end{center}
\caption{Rapidity distribution of Higgs bosons at 
LHC\protect\cite{dixhiggs}.}
\label{fig:hrap}
\begin{center}
\epsfig{file=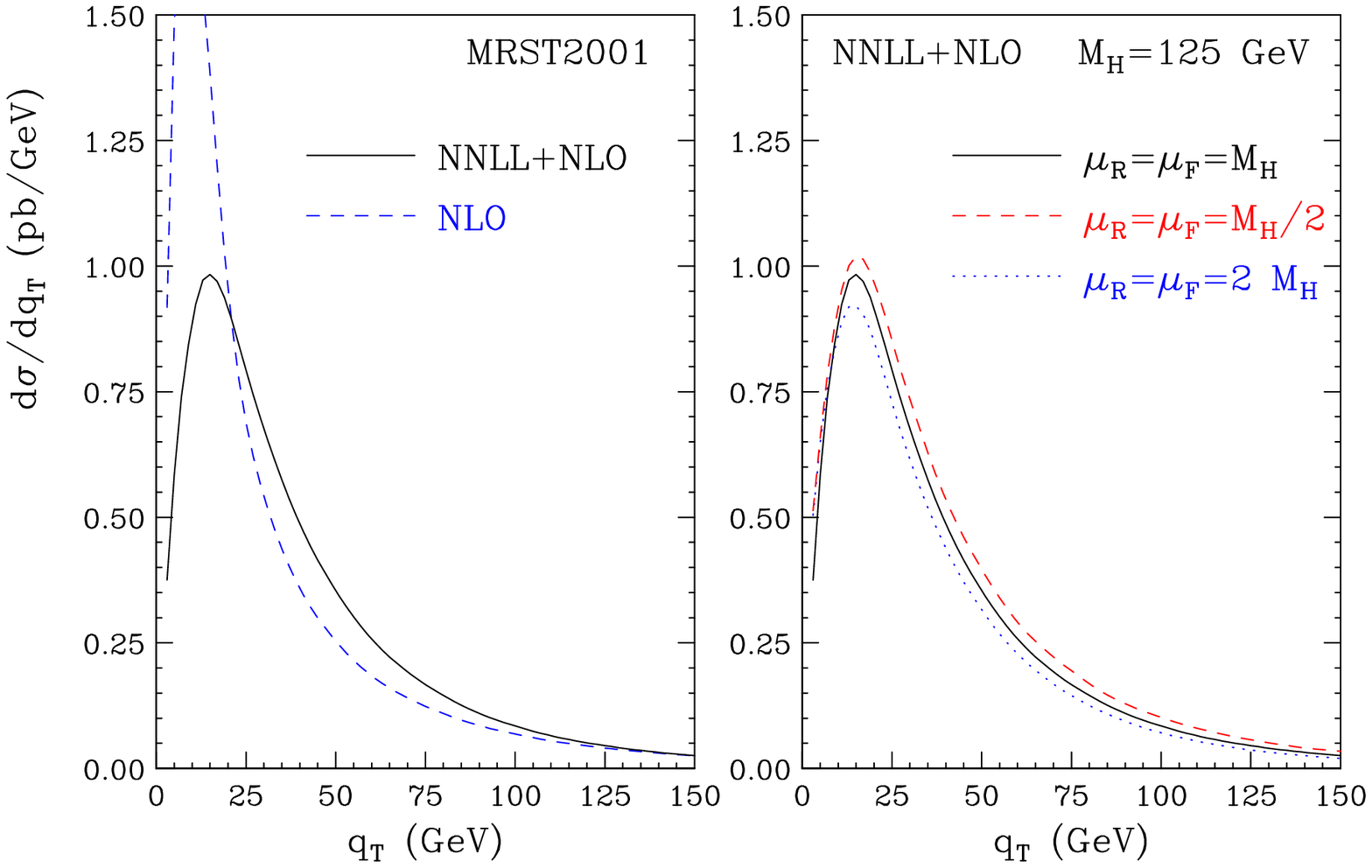,width=8cm}
\end{center}
\caption{Transverse momentum spectrum of Higgs bosons at 
LHC, including soft gluon resummation\protect\cite{cataniperp}.}
\label{fig:hqt}
\end{figure}

Since hadron collider experiments only cover a limited range of the 
final state phase space, it is very desirable to have not only 
predictions for the inclusive Higgs production cross sections, but also 
for differential distributions in rapidity and transverse momentum. 
Next-to-leading order corrections to both 
distributions\cite{dixhiggs,kunsztqt,vnqt} became available 
recently. The rapidity distribution, Figure~\ref{fig:hrap}, is only moderately 
modified, but extends out significantly beyond the experimental coverage. 
The calculation of corrections to transverse momentum distributions is 
reliable only for transverse momenta larger or equal to the Higgs boson 
mass, while multiple soft gluon radiation plays a crucial role at
lower transverse momenta. The resummation of these corrections 
was performed recently\cite{cataniperp} 
to NNLL accuracy, Figure~\ref{fig:hqt}. 

\subsection{Gauge Bosons}
The production cross sections for $W^\pm$ and $Z^0$ bosons at hadron colliders 
are well understood both experimentally and theoretically. At present,
these cross sections are measured to an error of about 10\% from 
Tevatron Run I, largely limited by statistics. A considerable reduction of the 
experimental error is anticipated from Run II and for the LHC. On the theory
side, inclusive vector boson production has been computed to 
NNLO\cite{dynnlo,hkhiggs}.

Given the good theoretical and experimental understanding of 
$W^\pm$ and $Z^0$ boson production, it has been suggested to use these 
for a determination of the LHC luminosity\cite{dittmar}. In practice, it 
turns out that it is not possible to measure the fully inclusive production 
cross sections, but only cross sections integrated over a  restricted range
in rapidity, for which NNLO corrections were also computed very 
recently\cite{dixdy}. For the $W^\pm$ 
production, which is observed only through the $l\nu$ decay channel, it 
is moreover mandatory to compute the spatial distribution of the decay 
products, which is however only known to NLO\cite{dyrad}. 

A crucial ingredient to precision NNLO
predictions of these cross sections 
at LHC are parton distributions accurate to this order. The determination 
of parton distributions from a global fit to experimental data is 
described in great detail in Robert Thorne's talk at this 
conference\cite{thorne}. To perform a fit at NNLO, one needs 
on the one hand the coefficient 
functions for all contributing observables (deep inelastic scattering,
Drell-Yan process, jet production and possibly direct photon production) 
to NNLO. At present, only Drell-Yan process and deep inelastic scattering 
are known to this order.
On the other hand, also the partonic splitting kernels 
(Altarelli-Parisi splitting functions) are required to NNLO accuracy. 
At the moment, this calculation is ongoing. The method applied in 
this calculation is the determination of the splitting functions 
from the forward photon-parton scattering amplitude at three loops, evaluated 
in moments of the external partonic momentum. Intermediate results 
involve some fixed 
moments\cite{apmom}, as well as all moments for the non-singlet 
fermionic loops\cite{apnf}.

\section{Conclusions and Outlook}
\label{sec:conc}
QCD affects all observables studied at present and future hadron colliders.
Given the anticipated luminosity of these machines, QCD reactions there 
 will be precision physics, 
very much like electroweak physics was precision physics in the LEP era. 
The study of many of the standard scattering reactions will allow a precise
determination of the strong coupling constant, electroweak parameters, 
quark masses and parton distribution functions. In turn, this information 
translates in improved predictions for new physics signals and their 
backgrounds. 

Both the precision determination of standard model parameters (and auxiliary 
quantities) and the design of search strategies for new physics effects 
require substantial input from theoretical calculations. For the purpose 
of drafting searches for physics beyond the standard model, one often 
requires predictions for multiparticle final states, which are at present 
only available at the leading order. Such accuracy is 
in general sufficient for this purpose, given that the available generic
programs also provide interfaces to partonic showers and hadronization models,
thus predicting fully hadronic events, which can be further processed 
through detector simulations. Next-to-leading order calculations 
will be important to refine searches, and to identify potential new
signals, since quantitative predictions start to become reliable only at
this order. The current frontier of  NLO calculations are $2\to 3$ reactions,
where some of the most prominent observables are known. Extension to 
$2\to 4$ reactions will require new theoretical tools, in particular towards 
generic, process independent algorithms. Another important development 
in NLO calculations is the interface to partonic showers, which has 
recently been devised. For the precise extraction of standard model parameters 
from benchmark reactions, NNLO calculations will be mandatory, since 
these reactions are (already at present colliders) measured to a level of 
accuracy at which the theoretical error on the NLO calculation becomes 
the dominant source of uncertainty. Presently, first NNLO calculations were 
performed for $2\to 1$ reactions, and $2\to 2$ calculations are well under way.
For applications at hadron colliders, NNLO calculations become only meaningful 
if augmented by parton distributions accurate to NNLO, which require the 
knowledge on the three-loop splitting functions, which are also calculated at 
present.

Many observables do moreover require the 
resummation of large logarithms spoiling the convergence of the perturbative 
series.
Fragmentation effects enter 
many observables with identified particles in the final state. In particular,
a consistent treatment of heavy quark fragmentation effects can account for 
a large part of the observed discrepancy in $B$ hadron spectra, and 
quark-to-photon as well as quark-to-pion fragmentation yield important 
contributions to photon pair final states forming an important
 background to Higgs searches. Much valuable information on these 
fragmentation functions is contained in data from LEP, and 
should be extracted (as long as this is still a feasible task)
to improve predictions for collider observables.

\vspace{15cm}

\vfill

\section*{DISCUSSION}

\begin{description}
\item[Sungwon Lee] (Texas A \& M University):
Currently there are large
discrepancies between data and NLO QCD predictions for direct photon
production.  Has there been any theoretical progress on the intrinsic
(effective parton) $k_t$ issue for direct photon production?

\item[Thomas Gehrmann:]
Understanding of  $k_t$ effects is one 
motivation for doing NNLO calculations.  At NNLO, for the first time, we
start to fully model $k_t$ effects due to hard parton emission in the initial
state because we allow for either double emission of one of the incoming
legs, or for double uncorrelated emission.  So, once you have
NNLO calculations for $2\to 2$ scattering processes available, you will really
have a theoretical tool for computing $k_t$ effects from perturbation theory.

\item[Ikaros Bigi] (Notre Dame University):
You showed this very instructive curve
about $b$ quark fragmentation, where you showed that the Peterson et al.\
prediction is less than optimal.  Do you have (or does someone have) a
similar curve for charm fragmentation?

\item[Thomas Gehrmann:]
The recent work by Cacciari and
Nason addresses this issue. They 
 refit charm fragmentation functions 
using a new parameterization in moment space trying to expose
the information content of LEP data 
relevant  to proton-antiproton 
colliders.  Comparison with old parameterizations 
is however not made. The basic message 
is that when you start fitting fragmentation functions, you have to be
extremely careful that you are not introducing artifacts from the choice
of parameterization which then, although giving a you a least $\chi^2$
fit, do not really reproduce the data because you are starting with
too stiff an initial form.

\end{description}

\end{document}